\documentclass[%
reprint,
superscriptaddress,
 amsmath,amssymb,
longbibliography,
]{revtex4-2}

\usepackage{romannum}
\usepackage{graphicx}
\usepackage{dcolumn}
\usepackage{bm}

\usepackage{siunitx}

\begin{document}

\title{\large{Nonlinear force response of modular lattice-based metamaterials}}

\author{Jochem G. Meijer}
\email[]{jgmeijer@uchicago.edu}
\affiliation{James Franck Institute, The University of Chicago, Chicago, IL 60637}

\author{Armin Youseﬁ}
\affiliation{Department of Mechanical Engineering, University of Colorado, Boulder, CO, 80309}

\author{Francois Barthelat}
\affiliation{Department of Mechanical Engineering, University of Colorado, Boulder, CO, 80309}

\author{Heinrich M. Jaeger}
\email[]{h-jaeger@uchicago.edu}
\affiliation{James Franck Institute, The University of Chicago, Chicago, IL 60637}
\affiliation{Department of Physics, The University of Chicago, Chicago, IL 60637}

\begin{abstract}
Lattice-based metamaterials provide lightweight platforms where local instabilities can govern the global mechanical response, enabling applications in energy routing, vibration isolation, and impact mitigation. 
Although much progress has been made in controlling deformation and buckling sequences through geometric design, the behavior of coupled nonlinear units over a large range of strain rates and their history-dependent response is less explored.
Here, we investigate lattice-based mechanical metamaterials whose nonlinear buckling behavior can be harnessed through modular architectures. 
By combining modular units in series, we show that their interaction gives rise to emergent force responses, including transient weakening and enhanced force attenuation, that are absent in the individual modules. 
Furthermore, selected designs exhibit training behavior under cyclic loading, transitioning between distinct buckling states and revealing a history-dependent mechanical response. 
Our results demonstrate that modular, instability-driven metamaterials can be programmed and tuned not only through geometry but also through loading history, opening new avenues for designing a nonlinear stress-response in mechanical systems. \\
\end{abstract}

\date{\today}

\maketitle

\section{Introduction}

Mechanical metamaterials that exploit Euler buckling display a wide range of striking behaviors, including auxetic behaviour \cite{mullin2007pattern, bertoldi2008mechanics, lakes1987foam, grima2000auxetic, grima2005auxetic, gaspar2005novel,wang2009hybrid, bertoldi2010negative,  greaves2011poisson, nicolaou2012mechanical,reid2018auxetic}, as well as controlled sequential deformations \cite{florijn2014programmable, shan2015multistable, frenzel2016tailored, coulais2018multi, meng2020multi}.
Such mechanical responses to applied strain or stress are particularly relevant for applications ranging from shielding and impact mitigation \cite{shan2015multistable, frenzel2016tailored, liu2024harnessing}, as well as vibration damping \cite{dykstra2023buckling}.
Sequential buckling in lattice-based metamaterials typically arises from a competition between local softening and subsequent stiffening. 
The softening stage is commonly associated with a negative stiffness, emerging at the onset of instability.
This localized deformation is then arrested by densification mechanisms of the lattice structure.

The sequence of instabilities can be tuned through the geometric design of the metamaterial \cite{frenzel2016tailored, coulais2018multi,  meng2020multi, li2021programmable, meng2022deployable, dudek2022micro, liu2024harnessing, liu2025tuning} and even steered \cite{meijer2026diffusive}. In addition,  entangled systems \cite{murphy2017aleatory,dalaq2025strength, pezeshki2026combined}, granular crystals \cite{karuriya2023granular, karuriya2024fully, karuriya2024plastic} or modular metamaterials \cite{zhao2025modular, zheng2025rigid, naderi2025stiff} have proven to be a versatile  approach to  achieve improved material properties and a targeted mechanical response by design. The additional potential of enabling programming of the response  allows for another layer of control \cite{ wagner2019programmable, naderi2025stiff, zou2023magneto, hu2023engineering, meng2022deployable, zhang2021tailored}.
These developments suggest a broader paradigm: lattice-based metamaterials can be treated not merely as bulk continua with unusual effective moduli, but as assemblies of coupled, nonlinear units whose interactions and loading history shape the global force response.

Here, we experimentally investigate modular lattice-based metamaterial units under compression whose highly nonlinear, strain rate-dependent and dissipative buckling characteristics are enhanced through controlled modular architectures. 
We show that the temporal weakening associated with local buckling can be exploited to create force responses of the ensemble that emerge only from the coupled dynamics of multiple units.
We also investigate the mechanical response of such ensembles over a wide range of applied strain rates.
Beyond programmability through design, we further demonstrate that selected architectures exhibit training behavior under cyclic loading, transitioning between distinct buckling states as the number of loading cycles increases if the proper training protocol has been chosen. 
This evolution reveals an additional axis of mechanical tunability, namely history dependence, linking geometric design, nonlinear instability, and mechanical memory.
Our work aims to provide proof of principle experiments regarding the capability of this type of metamaterials, and it opens pathways towards programmable force profiles and adaptive load management.

\begin{figure*}[t!]
  \centerline{\includegraphics[width=\textwidth]{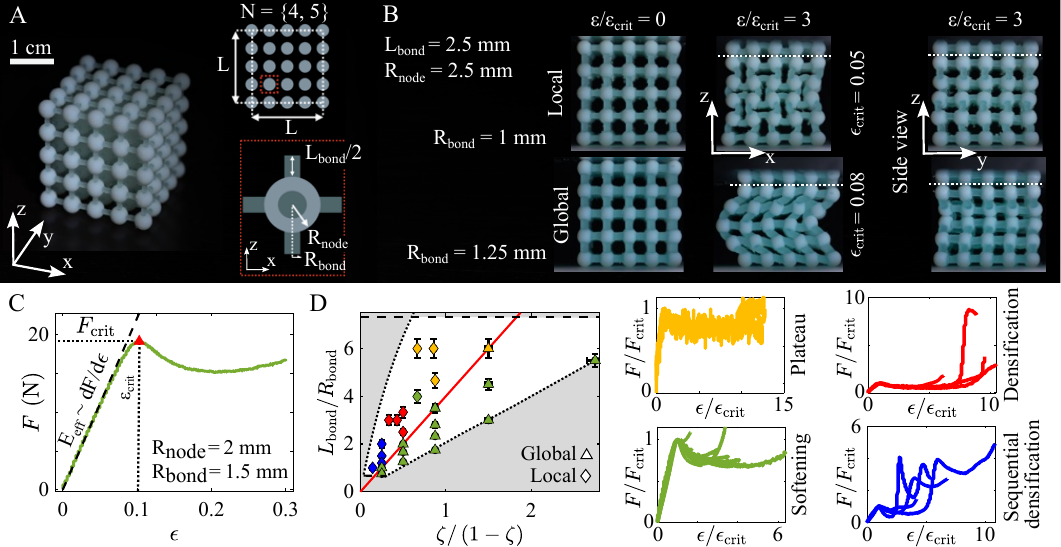}}
\caption{\textbf{Force response of 3D-printed cubic lattices under compression.}
(\textit{A}) 3D-printed lattice structure containing rigid nodes (white) and flexible bonds (lightblue) where $N = 5$, $L = \SI{30}{\mm}$, $R_{\mathrm{node}} = \SI{2.5}{\mm}$, $R_{\mathrm{bond}} = \SI{0.75}{\mm}$ and $L_{\mathrm{bond}} = \SI{2.5}{\mm}$. 
(\textit{B}) Experimental snapshots (front view) showing the two different buckling modes (local versus global) under uniaxial compression of two slightly different designs, triggered once a critical strain, $\epsilon_{\mathrm{crit}}$, is reached (see Suppl.\,Movie\,1). 
The rightmost panel shows the difference in self-folding of the five layers from the side. 
The white dotted lines indicate where the bottom of the compression plate touches the top of the lattice. 
(\textit{C}) Reaction force $F$ during uniaxial compression as a function of strain $\epsilon$ for the lattice design with $R_{\mathrm{node}} = \SI{2}{\mm}$ and $R_{\mathrm{bond}} = \SI{1.5}{\mm}$. 
The first local maximum denotes $F_{\mathrm{crit}}$ and $\epsilon_{\mathrm{crit}}$ (red triangle).
The initial slope (dashed line) gives the effective Young's modulus $E_{\mathrm{eff}} \sim \mathrm{d}F/\mathrm{d}\epsilon$ of the lattice.
(\textit{D}) Parameter space indicating the regions of global (triangles) and local (diamonds) buckling, according to Eq.\,\ref{Eq:criterion} \cite{meijer2026diffusive}. 
The gray area is beyond the available design space. 
Different symbol colors correspond to different characteristic force responses after buckling. 
The plots on the right show the normalised force profiles for the four different classifications, i.e., plateau (yellow), softening (green), densification (red), and sequential densification (blue).
}
\label{fig:1}
\end{figure*}

After introducing the experimental protocols and design specifications (Sec.\,\ref{sec:exp}), we focus on the dynamic and force response of the individual lattice-based metamaterial units we use as building blocks, and show how these can be tuned through architecture alone (Sec.\,\ref{sec:results_a}).
This is followed by our demonstration that a particular subset of the lattice-based metamaterials is susceptible to mechanical training, evolving throughout repeated cyclic compression while exhibiting mechanical memory .
We then discuss different coupling mechanism between individual units and investigate how this affects their strain rate-dependent mechanical properties, before addressing in more detail their potential to achieve a more targeted force response through their combination and mutual interactions (Sec.\,\ref{sec:results_b}).
We end with a conclusion in Sec.\,\ref{sec:conclusion}.

\section{Experiments}
\label{sec:exp}

The lattice-based metamaterials are fabricated using a high-resolution polyjet 3D-printer capable of multi-material deposition (Stratasys J850) and consist of rigid nodes (VeroWhite Polyjet Resin) connected by flexible bonds (Agilus 30 Polyjet Resin, Shore-A40), see Fig.\,\ref{fig:1}\textit{A} \cite{meijer2026diffusive}.
Unless mentioned otherwise, we restrict ourselves to $N = \{4, 5\}$ number of nodes per side (hence 64 or 125 nodes in total), inner lattice size $\SI{18}{\mm} \leq L \leq \SI{33.75}{\mm}$, node radius $\SI{1}{\mm} \leq R_{\mathrm{node}} \leq \SI{3.35}{\mm}$ and bond radius $\SI{0.75}{\mm} \leq R_{\mathrm{bond}} \leq \SI{2}{\mm}$. 
The bond length connects to the bond radius and the total size of the structure through $L_{\mathrm{bond}} = \left[ L - (2N-2)R_{\mathrm{node}} \right] / \left[ N-1 \right]$.
The printer resolution is $\pm \SI{50}{\micro \meter}$.

Uniaxial compression tests are performed on the modular units using three different experimental setups to achieve a range of strain rates that spans more than five order of magnitude. 
During compression the reaction force $F$ is measured as a function of the applied strain while the structure is observed by a side view camera, either a digital camera (Sony alpha 1) or high-speed camera (Vision Research Phantom and Photron FASTCAM Nova S20) to achieve high enough frame rates.

For the low strain rate experiments ($v_{\mathrm{crosshead}} = 0.03 - 3\, \SI{}{\mm \per \s}$) we use an universal materials tester (Instron 5600), straining the lattice to a certain target strain. 
The majority of the experiments are performed at a constant crosshead speed of $v = \SI{0.5}{\mm \per \s}$ and all compression tests are repeated at least three times.

To achieve moderate strain rates ($v_{\mathrm{crosshead}} = 10 - 100\, \SI{}{\mm \per \s}$) we use a different materials tester (ZwickRoell EZ001). 
All compression tests are again recorded and repeated at least three times.

Finally, high strain rate experiments ($v_{\mathrm{crosshead}} = 1 - 8.5\, \SI{}{\m \per \s}$) were performed using a gravity-driven impact tower (Industrial Physics Ray-Ran-FWT1-2000) modified with a 600 mm long acrylic (PMMA) bar mounted underneath the crosshead following a “Dropkinson bar” configuration \cite{Song2018Dropkinson} used to produce enough displacement to bring the sample to large deformations. Strain gauge stations were installed on the input and output bars to record impact forces.
Apart from the reaction force, measured at the top of the structure, this experimental setup also allows us to directly measure the force at its bottom.

\section{Results}

\subsection{Single units}
\label{sec:results_a}

\textbf{Dynamic response.} 
The dynamical response to uniaxial compression at $v_{\mathrm{crosshead}} = \SI{0.5}{\mm \per \s}$ of two slightly different lattice structures is shown as snapshots in Fig.\,\ref{fig:1}\textit{B}.
Making the bonds thicker while keeping the bond length and node size the same causes the overall response to transition from local buckling (upper row) to global buckling (lower row) once a critical strain, $\epsilon_{\mathrm{crit}}$, is reached \cite{meijer2026diffusive}.
During local buckling, the lattice collapses into itself due to counter-rotation of neighboring nodes, which results in an hourglass-like pattern when looked at from the 'front' (here the xz-plane).
In contrast, during global buckling the middle row shears out to one side and all nodes in the same row co-rotate.
While there are four possible rotation directions that can be triggered by the buckling instability, only two will persist, as proven analytically \cite{meijer2026diffusive}, and are visible in the images as clockwise and counter-clockwise rotation in the left panels in Fig.\,\ref{fig:1}\textit{B}, which we define as 'front view'.
The rightmost panel in Fig.\,\ref{fig:1}\textit{B} shows the corresponding  'side view,' highlighting the difference in self-folding of the five layers as well as the absence of node rotation in the yz-plane. 

As shown in our earlier work \cite{meijer2026diffusive} the transition from global buckling (facilitated by shear between lattice planes)  to local buckling (facilitated by bond bending) occurs when $4 k_b/(k_s L_{\mathrm{bond}}^2) = 1$. 
Here the bond bending stiffness $k_b = E_{\mathrm{bond}} I_{\mathrm{bond}}/L_{\mathrm{bond}}$ and the shear stiffness $k_s = \kappa G_{\mathrm{bond}} A_{\mathrm{bond}}/L_{\mathrm{bond}}$ depend on the effective Young's and shear moduli $E_{\mathrm{bond}}$ and $G_{\mathrm{bond}}$ of the lattice material. 
Assuming the lattice to be fabricated from rods as bonds and spheres as nodes yields $I_{\mathrm{bond}} = \pi R_{\mathrm{bond}}^4 / 4$ (bonds’ area moment of inertia), $A_{\mathrm{bond}} = \pi R_{\mathrm{bond}}^2$ (cross-sectional area) and $\kappa = \frac{1 - \zeta}{\zeta} \, R_{\mathrm{bond}}/L_{\mathrm{bond}}$ (shear correction factor) \cite{timoshenko2012theory, meijer2026diffusive}.
Here $\zeta = (N-1) L_{\mathrm{bond}}/L$ is the fraction of side length $L$ occupied by bonds. 
After substituting this for $k_b$ and $k_s$, the transition criterion becomes
\begin{equation}
\frac{L_{\mathrm{bond}}}{R_{\mathrm{bond}}} = \frac{E_{\mathrm{bond}}}{G_{\mathrm{bond}}} \frac{\zeta}{1 - \zeta}.
\label{Eq:criterion}
\end{equation}

\textbf{Force response.} 
For a lattice design with $N = 5$, $L = \SI{30}{\mm}$, $R_{\mathrm{node}} = \SI{2}{\mm}$ and $R_{\mathrm{bond}} = \SI{1.5}{\mm}$ the typical force-strain relationship is shown in Fig.\,\ref{fig:1}\textit{C}.
Initially linearly increasing with applied strain, $F$ reaches a maximum before entering the post-buckling regime.
The slope of the initial increase relates to the effective Young's modulus of the cubic lattice as $E_{\mathrm{eff}} \sim \mathrm{d}F / \mathrm{d} \epsilon$. 
The reaction force and strain at the first local maximum, $F_{\mathrm{crit}}$ and $\epsilon_{\mathrm{crit}}$, indicate the critical force and strain at which either global or local buckling occurs.
In line with Eq.\,\ref{Eq:criterion}, distinct domains that produce global (triangles) and local (diamonds) buckling are delineated in a plot of bond slenderness, $L_{\mathrm{bond}}/R_{\mathrm{bond}}$ versus the bond-to-node length ratio, $\zeta/(1 - \zeta)$, see Fig.\,\ref{fig:1}\textit{D}.
The areas shown in gray fall outside the available design space due to either printer resolution limitations (dashed line) or design restrictions (dotted line).
For the lattice structures investigated, we find the ratio of the effective Young's and shear moduli of the lattice material to be $E_{\mathrm{bond}}/G_{\mathrm{bond}} = 4$ (red line) \cite{meijer2026diffusive}. 

\begin{figure}[b!]
  \centerline{\includegraphics[width=\columnwidth]{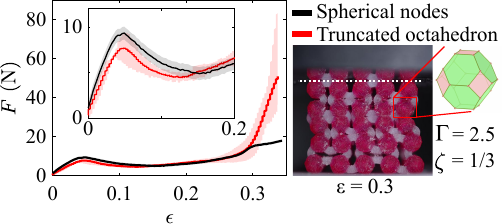}}
\caption{\textbf{Enhanced densification using non-spherical nodes.}
Reaction force $F$ during uniaxial compression as a function of strain for the same lattice designs with either spherical nodes (black) or truncated octahedra (red). Profiles are averaged over three experiments with the shaded regions indicating one standard deviation. Here, $\Gamma = L_{\mathrm{bond}}/R_{\mathrm{bond}}$.
}
\label{fig:2}
\end{figure}

The different force-strain relationships of all the lattice structures can be characterised into four different classifications, see right plots in Fig.\,\ref{fig:1}\textit{D}.
The different symbol colors correspond to different characteristic force responses after buckling, which are i) plateauing (yellow), ii) (mainly) softening (green; $ F_{\mathrm{max}}/F_{\mathrm{crit}} < 2$), iii) densification (red; $ F_{\mathrm{max}}/F_{\mathrm{crit}} \geq 2$), and iv) sequential densification (blue).
Whereas for the former two cases the bonds/nodes are sufficiently long/small that after buckling the rigid nodes do not touch, decreasing/increasing the bond length/node size will lead to node-on-node contact after local buckling.
Once strained enough, it causes a sudden rise in the reaction force of the lattice structure.
When the bonds/nodes are made even shorter/larger the initial node-on-node contact gives rise to sudden slip between horizontal layers, corresponding to the significant weakening after the second local maximum observed in the blue force profiles in Fig.\,\ref{fig:1}\textit{D}.
This is followed by a second densification stage (if strained enough) where the rigid nodes are forced into a denser configuration and make contact once more.
Once the strain is released, the structures relax back to their original configuration.

To enhance the initial densification stage we must ensure that the node-on-node contact has a larger resistance to slip. 
This can be achieved by replacing the spherical nodes with truncated octahedra \cite{karuriya2023granular, karuriya2024fully, karuriya2024plastic}.
By exploiting the known counter-rotation of the nodes during local buckling, a denser packing is achieved where the flat surfaces with higher friction will eventually come into contact, leading to a repeatedly more pronounced densification stage, see Fig.\,\ref{fig:2}. 
For this type of design the bonds connect to the non-spherical nodes through the squared faces of the truncated octahedron (red areas indicated in the schematic on the right of Fig.\,\ref{fig:2}).

\begin{figure}[t!]
  \centerline{\includegraphics[width=\columnwidth]{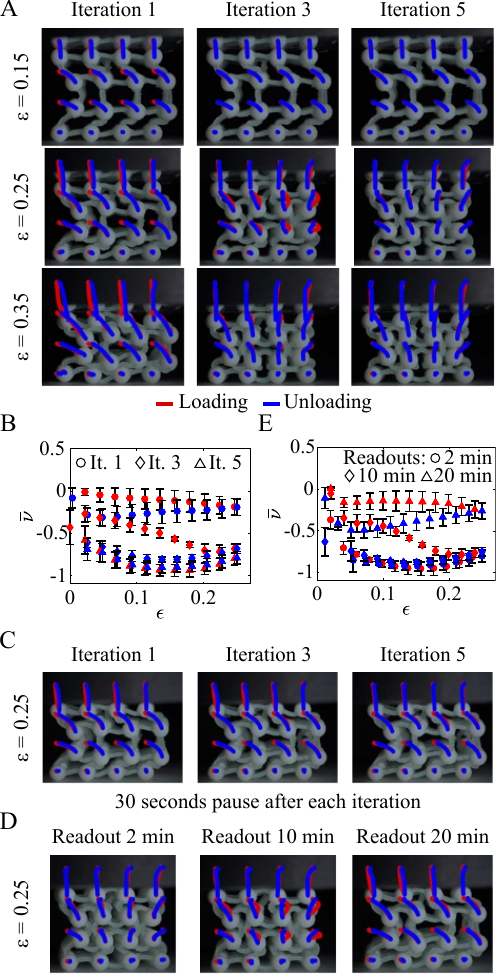}}
\caption{\textbf{Trainability of lattice-based metamaterials.}
(\textit{A}) Snapshots showing the dynamical response during cyclic loading for different target strains at for different iterations (see Suppl.\,Movie\,2). 
The red and blue traces show the compression (red) and decompression (blue) trajectories of the nodes for the given iteration.    
(\textit{B}) Poisson's ratio averaged over the center four nodes $\bar{\nu}$ as a function of strain. 
Different symbols correspond to different iterations and colors represent the (de)compression stage. 
(\textit{C}\,\&\,\textit{D}) Readouts of the dynamical response at the times indicated after the structure experienced cyclic loading to $\epsilon = 0.25$ without pauses in between the ten iterations.
(\textit{E}) Experimental snapshots during cyclic loading with $\SI{30}{\second}$ pauses after each iteration (see Suppl.\,Movie\,3).
}
\label{fig:3}
\end{figure}

\textbf{Trainability.} Thus far, the focus has been on the dynamic and force response during a single loading period. 
Yet, the polymeric nature of the polyjet-printed resin gives rise to viscoelastic behaviour of the bonds that - in combination with a well chosen design - should, in principle, make it possible to enter regimes where the dynamical response of the lattice-based metamaterial becomes trainable \cite{jaeger2024training}.
In such a case, repeated forcing via an appropriately chosen training protocol could condition the structure to respond in a targeted manner \cite{gowen2025training,gowen2025training2}.

In Fig.\,\ref{fig:3}\textit{A} we show how a lattice unit with $N = 4$, $L = \SI{29.3}{\mm}$, $R_{\mathrm{node}} = \SI{2.6}{\mm}$, $R_{\mathrm{bond}} = \SI{0.75}{\mm}$ and $L_{\mathrm{bond}} = \SI{4.5}{\mm}$ responds to such training.
By performing repeated compression cycles to a fixed target strain at $v_{\mathrm{crosshead}} = \SI{0.5}{\mm \per \s}$ with no pauses after each iteration, the magnitude of the target strain dictates if - and after how many iterations - the dynamic response of the lattice unit transitions from global to local buckling (see Fig.\,\ref{fig:3}\textit{A}).
The red and blue traces in the snapshots show the trajectories of the nodes for the given iteration during compression (red) and decompression (blue), respectively.    
Apart from highlighting the reversibility in the node trajectories during most of the iterations, they clearly highlight the transition from one buckling state (global) to the other (local), see middle panel. 
We quantify the degree to which the lattice can fold into itself under compression by determining the Poissons's ratio, $\nu = - \epsilon_{x} / \epsilon_{z}$, where $\epsilon_{x}$ and $\epsilon_{z}$ are the strain in the transverse and axial direction, respectively. 
Using the tracked positions of the nodes for a target strain of $\epsilon = 0.25$, we can determine the instantaneous Poisson's ratio $\nu_{i,k}(\epsilon)$ of every node and take an average of the four in the center.
Figure\,\ref{fig:3}\textit{B} shows the evolution of this average, $\bar{\nu}$, as a function of strain during compression (red) and decompression (blue) for different iterations, indicated by different symbols.
We transition from global buckling (circles) with $\bar{\nu} \approx 0$ to local buckling (diamonds and triangles) with $\bar{\nu} < 0$, and turn more auxetic as the number of iteration increases. 
Repeating the above-mentioned experiments but introducing a pause after each iteration allows for an additional control to prevent or delay the transition in the dynamic response.
We find that a pause of $\SI{30}{\second}$ in between cycles is already enough to delay the transition beyond five iterations when choosing a target strain of $\epsilon = 0.25$, see Fig.\,\ref{fig:3}\textit{C}.
It is worth noticing, that the characteristic polymer relaxation time is found to be approximately one second (see Appendix\,C).

Finally, performing a cyclic training protocol of ten iterations to a target strain of $\epsilon = 0.25$ without pauses between iterations, then allows us to determine how long the lattice-based metamaterial 'remembers' its training.
Experimental snapshots and the corresponding quantification through $\bar{\nu}$, shown in Figs.\,\ref{fig:3}\textit{D}\&\textit{E}, indicate that only after more than ten minutes the structure has transitioned back towards its prior (global) buckling mode.

\subsection{Coupled units}
\label{sec:results_b}

\textbf{Coupling mechanism.} 
The design of the lattice-based metamaterials thus dictates their force-strain relationship and the structures can essentially be seen as highly non-linear, viscoelastic springs with strain- and strain-rate dependent stiffness (see Appendix\,A).
Given the symmetric nature of the cubic lattices, individual units can be combined so their dynamical and mechanical responses become coupled. 

To highlight how the coupling mechanism changes this response, two identical units are placed in series with either a very stiff connection (stacked) or a flexible one (elongated).
For the former, a thin glass plate of $\SI{2}{\mm}$ thickness separates the two structures and functions as a stiff connection. 
For the latter, the structure is fabricated in such a way that the individual units are connected through an additional layer of flexible bonds.  
The individual units' designs are characterised by bond slenderness, $\Gamma = L_{\mathrm{bond}}/R_{\mathrm{bond}}$, and the fraction of side length $L$ occupied by bonds, $\zeta$. 
Here, $\Gamma = 2.5$, $\zeta = 1/3$.

\begin{figure}[h!]
  \centerline{\includegraphics[width=\columnwidth]{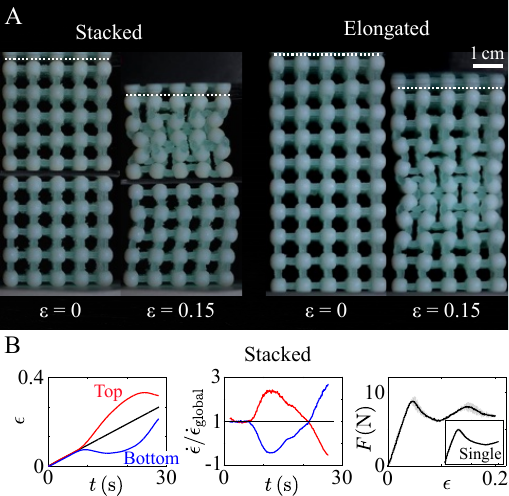}}
\caption{\textbf{Stacked versus elongated.}
(\textit{A}) Experimental snapshots of two identical units ($\Gamma = 2.5$, $\zeta = 1/3$) in series under compression ($v_{\mathrm{crosshead}} = \SI{0.5}{\mm \per \s}$), see Suppl. Movie\,4.
For the stacked case (left), the units are connected by a thin glass plate separating the two.
For the elongated case (right), individual units are connected through an additional layer of bonds. 
(\textit{B}) Strain (left) and strain rate (middle) as a function of time for the top (red) and bottom (blue) unit, respectively. The black lines indicate the globally applied strain at constant strain rate. Reaction force (right) as a function of global strain, averaged over three experiments with the shaded regions indicating one standard deviation.
The inset shows the force response of the single unit. }
\label{fig:4}
\end{figure}

In Fig.\,\ref{fig:4}\textit{A} experimental snapshots are shown for both configuration during uniaxial compression.
Whereas for the elongated structure (right) local buckling nucleates at the middle, the individual units of the stacked configuration (left) experience their local buckling event sequentially (also Suppl. Movie\,4).  

When quantifying the local strain of the top and bottom units of the stacked system (see Fig.\,\ref{fig:4}\textit{B}, left) it becomes evident that, while initially both unit are strained identically, upon reaching $\epsilon = \epsilon_{\mathrm{crit}}$
only one will undergo buckling.
This will affect the local strain experienced by the other due to the pronounced weakening during post-buckling. 
Specifically, the top unit (red line in Fig.\,\ref{fig:4}\textit{B}, left) reaches far larger local strain than the bottom one (blue line Fig.\,\ref{fig:4}\textit{B}, left).
Correspondingly, the experienced local strain is not constant (see Fig.\,\ref{fig:4}\textit{B}, middle) and either exceeds (top) or is below (bottom) the applied global strain rate.
The local strain rate of the bottom unit even becomes negative, meaning this unit is temporally un-strained and can even relax.
Eventually, when the global strain is increased, the second unit will buckle too.

This drastic change in the dynamic response also alters the force response of the coupled system.
Whereas the elongated unit shows a similar force-strain relationship to the single unit, the stacked configurations shows a second stiffening, followed by a second weakening (see Fig.\,\ref{fig:4}\textit{B}, right).
The coupling and modular force response of \emph{different} units is discussed later.

\textbf{Strain rate-dependent stiffness and strength.} 
As addressed in the previous section, the mechanical response of the lattice-based metamaterials is thought to be strain rate-dependent.
To investigate this more thoroughly, uniaxial compression tests are performed on the single, stacked and elongated structure, with $\Gamma = 2.5$, $\zeta = 1/3$.
The strain rate is varied over more than five orders of magnitude. 

\begin{figure}[t!]
  \centerline{\includegraphics[width=\columnwidth]{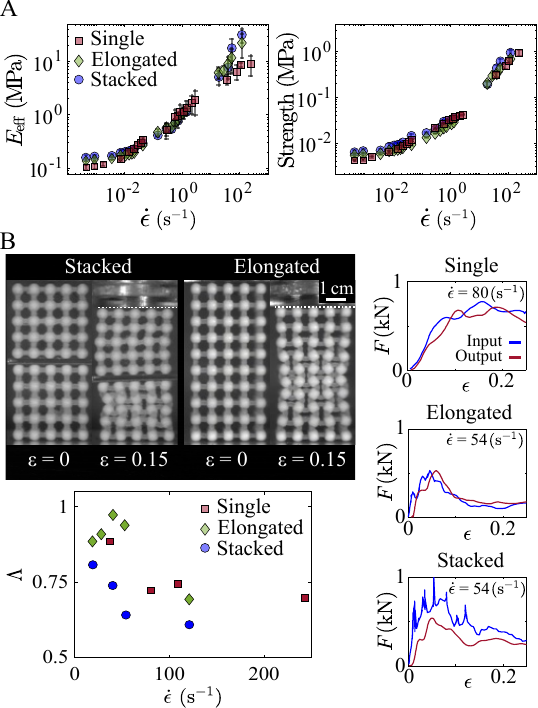}}
\caption{\textbf{Stiffness, strength and force attenuation.}
(\textit{A}) Effective Young's modulus $E_{\mathrm{eff}} = \left(\mathrm{d}F/\mathrm{d}\epsilon \right)/A_{\mathrm{cross}}$ of the lattice (left) and strength - characterised by $F_{\mathrm{crit}}/A_{\mathrm{cross}}$ - (right) as a function of strain rate.
(\textit{B}) Experimental snapshots of two identical units ($\Gamma = 2.5$, $\zeta = 1/3$) in series under impact ($v_{\mathrm{crosshead}} = \SI{3.75}{\m \per \s}$), see Suppl. Movie\,5.
Input (blue) and output (red) forces measured at the top and bottom of the structure, respectively, as a function of strain (right).
Force attenuation $\Lambda = \int_{0}^{\epsilon = 0.2} F_{\mathrm{out}} / \int_{0}^{\epsilon = 0.2} F_{\mathrm{in}}$ for the different designs during impact as a function of strain rate (bottom left).}
\label{fig:5}
\end{figure}

As indicated in Fig.\,\ref{fig:1}\textit{C}, the effective Young's modulus of the structure, $E_{\mathrm{eff}}$, is determined through the initial linear slope of the force-strain curve divided by the cross sectional area of the structure, $A_{\mathrm{cross}}$. 
Similarly, the strength of the structure is characterised by $F_{\mathrm{crit}}/A_{\mathrm{cross}}$. 
Figure\,\ref{fig:5}\textit{A} shows $E_{\mathrm{eff}}$ and the strength as a function of strain rate for the three different configurations. 
For the low and moderate strain rate experiments the scatter resembles an average over at least three experiments.
For the high strain rate experiments it resembles the measurement error within a single trial.

Whereas the strength seems indifferent on how the units are coupled, the stiffness does vary, especially at high strain rates.
The low strain rate plateau indicates a quasi-static regime in which the effective Young's modulus is independent of loading rate. 
Above a critical strain rate, viscoelastic effects become significant, leading to an increase in stiffness with increasing strain rate, following a power-law behaviour, with $E_{\mathrm{eff}} \propto \dot{\epsilon}^n$.
Our data suggests that the coupled units increase more rapidly in stiffness as compared to the single unit at higher strain rates.

\textbf{Force attenuation during impact.} 
During the high-strain-rate experiments, we also measured the force at the bottom of the lattice structure, which we refer to as the output force.
This allows us to investigate the force attenuation during the initial stages of impact. 
The experimental snapshots in Fig.\,\ref{fig:5}\textit{B} show impact at $v_{\mathrm{crosshead}} = \SI{3.75}{\m \per \s}$ for a stacked and elongated configuration. 
Although the structures are strained at much larger rates, their dynamic response is identical to that reported in Fig.\,\ref{fig:4}\textit{A}, albeit that in the stacked case the order in which the individual units buckle is reversed. 
This is not surprising, since both units have the same $\epsilon_{\mathrm{crit}}$ and small imperfections will biases the order. 

The plots on the right in Fig.\,\ref{fig:5}\textit{B} show representative force-strain curves for the  input and output forces measured for each of the three designs.
The force attenuation, $\Lambda = \int_{0}^{\epsilon = 0.2} F_{\mathrm{out}} / \int_{0}^{\epsilon = 0.2} F_{\mathrm{in}}$, is found to depend on strain rate as shown in the bottom left plot of Fig.\,\ref{fig:5}\textit{B}. 
At the lowest strain rates $\Lambda$ is approaching unity, meaning the applied load is being transmitted  through the  structure with little to no loss in the quasi-static limit.
With increasing strain rate the value of $\Lambda$ decreases to levels around 0.6 - 0.7, indicating significant attenuation of the transmitted load.
Of the designs tested, the stacked configuration reaches the lowest $\Lambda$ value, indicating potential for impact-mitigating applications. 

\begin{figure}[t!]
  \centerline{\includegraphics[width=\columnwidth]{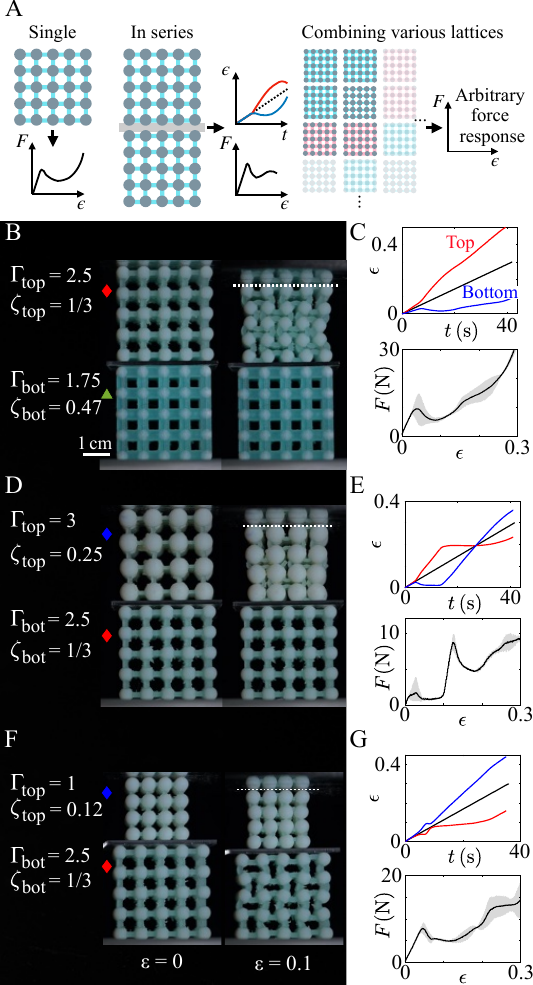}}
\caption{\textbf{Modular force response of lattices in series.}
(\textit{A}) Conceptual sketches of various lattice combinations to achieve a wide range of different force responses, by altering lattice designs and potentially material properties (red structures).
(\textit{B}) Experimental snapshots of two different lattice structures in series under compression at $v_{\mathrm{crosshead}} = \SI{0.5}{\mm \per \s}$, connected by a thin glass plate separating the two (see Suppl.\,Movie\,6). 
(\textit{C}) Upper: Local (red/blue for top/bottom) and global (black) strain as a function of time.
Lower: Reaction force $F$ during uniaxial compression as a function of global strain for the corresponding lattice structures in series, averaged over three experiments with the shaded regions indicating one standard deviation.
(\textit{D}-\textit{G}) Similar to \textit{A}\,\&\,\textit{B} for different combinations (see Suppl.\,Movies\,7\,\&\,8).
}
\label{fig:6}
\end{figure}

\textbf{Modular force response.} 
We can extend the control over the overall force-strain relationship of an ensemble of such modular units by combining \emph{different} individual units in parallel or series (or both), see sketch in Fig.\,\ref{fig:6}\textit{A}.

To show this conceptually, we place various designs with different individual characteristic force-strain relationships in series through stacking them.
The individual units' designs are again characterised by bond slenderness, $\Gamma = L_{\mathrm{bond}}/R_{\mathrm{bond}}$, and the fraction of side length $L$ occupied by bonds, $\zeta$, as well as a colored symbol corresponding to Fig.\,\ref{fig:1}\textit{D} indicating their typical force response after buckling.

Experimental snapshots of the lattice-based metamaterial in series are shown in Figs.\,\ref{fig:6}\textit{B},\,\textit{D},\,\&\,\textit{F}. 
Depending on its value of $\epsilon_{\mathrm{crit}}$ one unit might undergo buckling, which will affect the local strain experienced by the other due to the pronounced weakening during post-buckling. 
To highlight this interplay, we determine the local strains experienced by the top (red) and bottom (blue) unit and show them as a function of time in the upper plots in Figs.\,\ref{fig:6}\textit{C},\,\textit{E},\,\&\,\textit{G}.
Depending on the chosen combination, the experienced local strain might be larger than the overall applied global strain.
In addition, whereas globally the strain rate remains constant throughout the compression, the locally experienced strain rates vary significantly (see Appendix\,B).
They can even become zero or negative, meaning one of the structures is temporally un-strained or can even relax.
During this (short) relaxation period energy is suddenly being dissipated by the viscoelastic bonds.
The non-trivial combination of the individual strain-, strain rate- and relaxation history dependent force responses then yields the overall measured force response of the lattice units in series, shown in the lower plots in Figs.\,\ref{fig:6}\textit{C},\,\textit{E},\,\&\,\textit{G}.
It allows for the tuning of a more gradual densification stage (Fig.\,\ref{fig:6}\textit{C}), the introduction of a prolonged period where $\mathrm{d}F/\mathrm{d}\epsilon \approx 0$ for the entire structure (Fig.\,\ref{fig:6}\textit{E}), and the ability to create sudden kinks in the force response due to rapid, consecutive buckling (Fig.\,\ref{fig:6}\textit{G}), that are particularly pronounced in the local strain profiles, and even more so in those of the local strain rate (see Appendix\,B).

\section{Conclusions}
\label{sec:conclusion}

Our results show that lattice-based metamaterials can function as versatile modular units for designing complex adaptive responses to applied mechanical load.  
Using as our units square lattices consisting of rigid nodes connected with flexible bonds, which can undergo different buckling modes under compression, we demonstrate experimentally how widely different and highly nonlinear load responses can be achieved when such units are combined in series. 
In addition, by performing experiments in a range of strain rates covering more than five orders of magnitude, we highlight how the configurations of individual units affect the overall strength and stiffness of the ensembles.
Lastly, for units that are 3D printed from viscoelastic resin, we also show how the type of buckling behavior, and thus the dynamic response, is trainable via a suitable conditioning protocol.

\section*{Conflicts of interest}
There are no conflicts to declare.

\section*{Acknowledgements}
This work was supported by the Army Research Office under award W911NF-24-2-0184.   The University of Chicago Materials Research Science and Engineering Center, which is supported by the National Science Foundation under award DMR-2011854, provided additional support for sample fabrication through its shared experimental facilities.
The impact testing setup was acquired through the Defense University Research Instrumentation Program (DURIP) under Award Number 82428-TE-RIP.

\section*{Data availability statement}
The materials underlying this study are provided in the Materials Data Facility public repository (DOI: \url{https://doi.org/10.18126/6m80-h514}). 
Additional details regarding the supplementary movies can be found in \cite{suppl}.

\section{Author contributions statement}
J.G.M., and H.M.J. designed research; J.G.M. and A.Y. performed experiments; J.G.M. and A.Y. analyzed data; All authors contributed to the writing of the paper.

\section*{Appendix A: Strain rate dependent force response}
\begin{figure}[t!]
  \centerline{\includegraphics[width=\columnwidth]{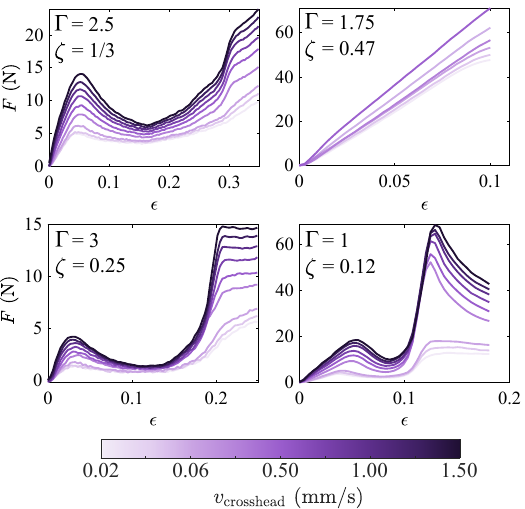}}
\caption{Force versus strain curves for different crosshead velocities of the lattice-based metamaterials corresponding to Fig.\,\ref{fig:6} of the main text. All curves show the average of three repetitions. }
\label{fig:7}
\end{figure}

The reaction force of the 3D-printed lattice-based metamaterial is strain rate dependent due to the viscoelastic behaviour of the flexible bonds. 
Figure\,\ref{fig:7} shows the profiles for the four designs used in Fig.\,6 of the main text at various crosshead velocities. 
Generally, the lattice structures are stiffer the higher the strain rate.

\section*{Appendix B: Strain rate dependent force response}
\begin{figure}[t!]
  \centerline{\includegraphics[width=\columnwidth]{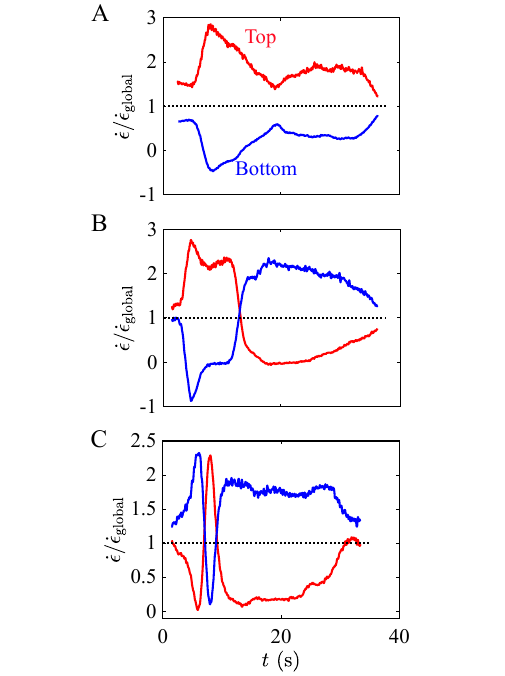}}
\caption{(\text{A}-\textit{C}) Local strain rate profiles normalised by the (constant) applied global strain rate ($\dot{\epsilon}_{\mathrm{global}}$) as a function of time corresponding to the curves shown in the upper plots of Fig.\,6\textit{B},\textit{D},\&\textit{F}, respectively.}
\label{fig:8}
\end{figure}

Figure\,\ref{fig:8} shows the local strain rate ($\dot{\epsilon}$) profiles corresponding to the curves shown in the upper plots of Fig.\,\ref{fig:6}\textit{B},\textit{D},\&\textit{F}.
Especially Fig.\,\ref{fig:8}\textit{C} highlights the effect of rapid, sequential buckling of lattices placed in series with similar values of critical strain, as mentioned in the main text.  

\section*{Appendix C: Polymer relaxation time }
\begin{figure}[t!]
  \centerline{\includegraphics[width=\columnwidth]{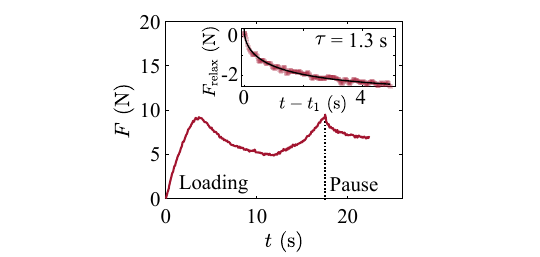}}
\caption{Force versus time curve of a lattice structure ($\Gamma = 2.5$ and $\zeta = 1/3$) being loaded to $\epsilon_{\mathrm{target}} = 0.25$ after which a $\SI{5}{\s}$ pause is initiated. The inset shows the force relaxation including the fitted Eq.\,\ref{eq:Frelax} (black line).  }
\label{fig:9}
\end{figure}

To obtain an estimate of the typical polymer relaxation time we perform the following experiment.
We compress the structure ($\Gamma = 2.5$ and $\zeta = 1/3$) to a target strain ($\epsilon_{\mathrm{target}} = 0.25$) at a constant crosshead velocity ($v_{\mathrm{crosshead}} = \SI{0.5}{\mm \per \s}$), after which we apply a $\SI{5}{\s}$ pause while the structure is being compressed. 
During this pause, we measure the temporal evolution of the reaction force, see Fig.\,\ref{fig:9}.
The pause is initiated at $t = t_1$ and the measured force relaxation (shown in the inset of Fig.\,\ref{fig:9}) is fitted by a stretched exponential:
\begin{equation}
F_{\mathrm{relax}} = \alpha \left( \mathrm{exp}\left[^-\left( \frac{t-t_1}{\tau} \right)^{\beta}\right] - 1\right).
\label{eq:Frelax}
\end{equation}
Here, the characteristic polymer relaxation time is denoted by $\tau = \SI{1.3}{\s}$.
For this case, $\alpha = \SI{3.3}{\N}$ and $\beta = 0.5$.

\bibliography{References}

%
%
%
%
%
%
%
%

\end{document}